\documentclass[layout=twocolumn,journal=amlccd,manuscript=article]{achemso}

\usepackage{graphicx}
\usepackage{amsmath}

\let\oldmaketitle\maketitle
\let\maketitle\relax

\title{Equilibration of High Molecular-Weight  Polymer Melts: A Hierarchical Strategy}

\author{Guojie Zhang}

\affiliation{Max Planck Institute for Polymer Research, Ackermannweg 10, 55128 Mainz, Germany}

\author{Livia A. Moreira}

\affiliation{Max Planck Institute for Polymer Research, Ackermannweg 10, 55128 Mainz, Germany}

\author{Torsten Stuehn}

\affiliation{Max Planck Institute for Polymer Research, Ackermannweg 10, 55128 Mainz, Germany}

\author{Kostas Ch. Daoulas}

\affiliation{Max Planck Institute for Polymer Research, Ackermannweg 10, 55128 Mainz, Germany}
\alsoaffiliation{Innovation Lab GmbH, Speyerer Stra{\ss}e 4, 69115 Heidelberg, Germany}

\author{Kurt Kremer}

\affiliation{Max Planck Institute for Polymer Research, Ackermannweg 10, 55128 Mainz, Germany}
\email{kremer@mpip-mainz.mpg.de}

\begin{document}
\twocolumn[
\begin{@twocolumnfalse}
\oldmaketitle
\begin{abstract}
A strategy is developed for generating equilibrated high molecular-weight polymer melts described
with microscopic detail by sequentially backmapping coarse-grained (CG) configurations.
The microscopic test model is generic but retains features like hard excluded volume interactions
and realistic melt densities. The microscopic representation is mapped onto a model of soft spheres with fluctuating size,
where each sphere represents a microscopic subchain with $N_{\rm b}$ monomers. By varying $N_{\rm b}$ a hierarchy of CG representations
at different resolutions is obtained. Within this hierarchy, CG configurations equilibrated with Monte Carlo at low resolution are sequentially
fine-grained into CG melts described with higher resolution. A Molecular Dynamics scheme is employed to slowly introduce the microscopic details
into the latter. All backmapping steps involve only local polymer relaxation thus the computational efficiency of the scheme is independent
of molecular weight, being just proportional to system size. To demonstrate the robustness of the approach, microscopic configurations 
containing up to $n=1000$ chains with polymerization degrees $N=2000$ are generated and equilibration is confirmed by monitoring key structural 
and conformational properties. The extension to much longer chains or branched polymers is straightforward.
\end{abstract}
\end{@twocolumnfalse}
]
Studying equilibrium and rheological properties of melts of long polymer chains with computer
simulations requires the preparation of equilibrated configurations described with microscopic detail. For this purpose, stochastic approaches 
have been proposed to circumvent the prohibitively large relaxation times in schemes with physically realistic dynamics, resulting from chain entanglements. 
Among methods addressing directly the microscopic scale, re-bridging (RB) algorithms~\cite{DBmoves} are the most advanced, modifying the chain connectivity while avoiding significant changes in local monomer packing. 
Even with their help, the longest melts currently addressed are those of linear polyethylene, corresponding to monodisperse samples with a few $C_{\rm 1000}$ chains~\cite{DBmoves}. 
Introducing polydispersity, increases the acceptance rate of RB moves and longer chains can be modeled. However, the system becomes 
less well-defined, e.g., for understanding rheological behavior and the samples remain rather small: the longest $C_{\rm 6000}$ (average length) melt~\cite{Uhlherr} that was realized contained $32$ chains. 
To prove equilibration these studies relied on the decay of conformational correlations. However, recent findings~\cite{wittmer2007} demonstrate
that the combination of chain connectivity and limited compressibility affects chain conformations. Since RB moves are 
largely decoupled from density fluctuations, such subtle effects suggest~\cite{wittmer2007} that to verify unambiguously melt equilibration more sensitive 
descriptors of chain shape, such as internal distance plots~\cite{wittmer2007, auhl2003}, should be considered.

To overcome the limitations encountered when modeling polymers directly at the microscopic scale, configuration-assembly procedures~\cite{auhl2003, Livia, HoeMuel, Brown}
have been proposed. Chains with conformations drawn from the distribution expected in the melt are treated as rigid bodies 
and arranged, under relaxed excluded volume interactions. The latter are reintroduced and the configuration is equilibrated. For relatively short chains, 
random placement of molecules has been sufficient~\cite{Brown} although this leads to unrealistically high density fluctuations: if $\rho_{\rm o}$ 
is the average melt monomer density, the isothermal compressibility, $\kappa_{\rm T}$, of the ideal gas of chains increases with polymerization degree $N$
as $\kappa_{\rm T} \sim N/\rho_{\rm o}$. For long polymers strong density fluctuations cause significant conformational distortions when introducing
excluded volume~\cite{auhl2003}. Density fluctuations can be reduced by optimizing molecular packing through a Monte Carlo (MC) scheme~\cite{auhl2003, Livia}.
Since the computational time of such an optimization increases significantly with chain length, such approaches are better
suited for melts with medium sized chains~\cite{Livia} ($N\!\sim\!500$). Postulating a distribution for polymer conformations is an additional limitation.
For example, due to the long-range intramolecular correlations~\cite{wittmer2007}, assuming ideal chain
statistics is an approximation. It is also difficult to estimate conformational properties in melts with non-linear molecules, close to spatial inhomogeneities, mixtures, etc.
\begin{figure}[h!]
\includegraphics[width=0.42\textwidth]{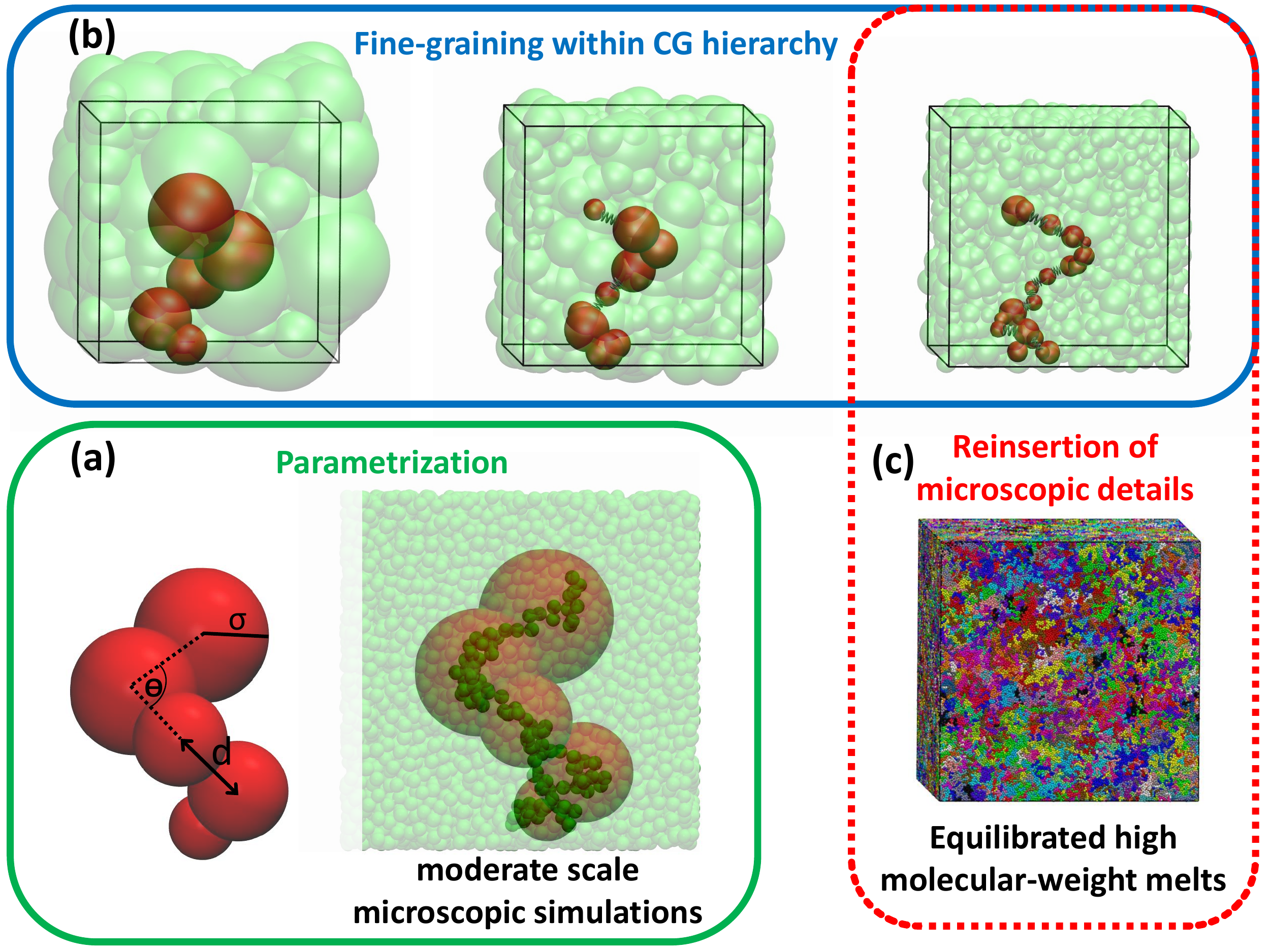}
\caption{Hierarchical backmapping strategy summarized. (a) A chain of soft spheres with fluctuating size is shown on the left (red) and on the
right (semi-transparent object) representing on CG level a polymer described with the microscopic KG
model (solid beads are monomers). Each sphere corresponds to a $N_{\rm b}$-monomer subchain. The CG force-field is
parameterized using data obtained from microscopic melts of medium sized chains, accessible to more conventional techniques. (b) Hierarchy of
sequentially fine-grained CG models where the resolution is doubled at each step. (c) Equilibrated high molecular-weight polymer melts are 
obtained, reinserting microscopic details into the last CG configuration of the fine-graining hierarchy. The snapshot shows a melt of $n=1000$ chains with polymerization
degree $N=2000$ ($2\times 10^6$ monomers in total). Colors are randomly chosen to improve visibility of different chains.}
\label{method}
\end{figure}

Universality is a key feature of polymeric systems which has tremendously facilitated their theoretical and experimental description: 
while on microscopic level their behavior depends strongly on chemical details, on mesoscopic scales it can be frequently related 
to a few generic parameters~\cite{DeGennesBook}. Benefiting from scale-separation, we develop a hierarchical backmapping strategy (Figure~\ref{method}) 
which opens the way for generating microscopic configurations of melts with chains significantly longer than those that can be addressed within 
the currently available techniques (more detailed discussion follows at the end of the Letter). Firstly, equilibrated configurations reproducing correctly 
the properties of the microscopic melt on length scales comparable to the size of the polymer coil are obtained, with minimum computational requirements, 
using a crude coarse-grained (CG) model based on a soft-blob description. We proceed to finer scales reducing the size of the blobs, through sequential 
fine-graining. Since the mesoscopic polymer structure is already captured, this procedure requires at each step only {\it local} relaxation of conformations 
and liquid structure, being thus computationally very efficient. Once the highest resolution in the hierarchy is reached, microscopic details can be efficiently reinserted, requiring again only local relaxation. 

Focusing on method development, we employ here the generic KG (Kremer-Grest) microscopic model~\cite{kremer1990} characterized, however, by hard excluded 
volume interactions and strong covalent bonds. Thus topological constraints hampering the equilibration of chemistry-specific atomistic models can be reproduced.
Each of the $n$ linear chains in the KG melt consists of $N$ monomers with mass $m$ linked by FENE springs. The non-bonded interactions 
of two monomers with distance $r$ are described through a purely repulsive Lennard-Jones (LJ) potential, $U_{\rm LJ}(r)$,
truncated at $r_{\rm c}$ and shifted so that $U_{\rm LJ}(r_{\rm c})=0$. In this work $k_BT = 1$, the LJ length, energy, 
and time scales are fixed to $\sigma_{\rm o} = 1$,  $\epsilon_{\rm o} = 1 $, and $\tau = 1$, while $m=1$. In these units the spring constant 
and the maximum extension of the bond are $k=30$ and $R_{\rm 0} = 1.5$, respectively, while $r_{\rm c} = 2^{1/6}$. The melt monomer density is $\rho_{\rm o} \simeq 0.85$. 
With the KG model, Molecular Dynamics (MD) simulations using the ESPResSo\verb!++! package~\cite{ESPR} were performed to generate 
reference data on the properties of melts with $N = 1000$. These demanding MD simulations will be denoted as {\it reference simulations}.

The KG model is coarse-grained by mapping each subchain with $N_{\rm b}$ monomers onto a soft sphere with fluctuating size~\cite{vettorel2010}. The coordinates of the 
center, ${\bf r}_{\rm i}$, and the radius, $\sigma_{\rm i}$, of the ${\rm i}$-th sphere match the position of the center-of-mass (COM), ${\bf R}_{\rm cm}$, and the instantaneous gyration 
radius, $R_{\rm g}$, of the ${\rm i}$-th subchain. Each polymer is represented by a chain of $N_{\rm CG} = N/N_{\rm b}$ spheres, as illustrated in Figure~\ref{method}(a). Varying $N_{\rm b}$, a 
hierarchy of models with different resolutions can be obtained (see Figure~\ref{method}(b)). The spheres are linked by bond, $\beta V_{\rm bond}(d)= 3d^{2}/2b_{\rm CG}^2$ , and 
angular~\cite{Suter}, $\beta V_{\rm bend}(\theta) = k_{\rm bend}(1+\cos \theta)/2$ potentials ($\beta = 1/k_BT$), where $d$ and $\theta$ stand for the distance and the angle, 
respectively, between consecutive spheres and bonds in a chain. The potentials $\beta V_{\rm sphere} (\sigma) = a_{\rm 1} N_{\rm b}^3\sigma^{-6} + a_{\rm 2} N_{\rm b}^{-1}\sigma^2$ 
and $\beta V_{\rm self}(\sigma) = a_{\rm 3}\sigma^{-3}$ are associated with each sphere radius, controlling its fluctuations. The former reproduces~\cite{Lhuillier1988} the distribution of $R_{\rm g}$ 
of ideal subchains (i.e., when all non-bonded interactions, apart from intramolecular 1-2, are set to zero). The latter accounts for subchain swelling by microscopic non-bonded 
interactions following Flory~\cite{Flory}. Non-bonded interactions between two spheres, ${\rm i}$ and ${\rm j}$, are given by $\beta V_{\rm nb}(r_{\rm ij}) =  \epsilon U_{\rm G}(r_{\rm ij})$ 
where $r_{\rm ij}$ is the distance of their centers, $U_{\rm G}(r_{\rm ij})$ is a Gaussian function with variance $\overline{\sigma^2} = \sigma_{\rm i}^2 + \sigma_{\rm j}^2$ (normalized to unity in three dimensions). 
The number of neighbors a sphere interacts with, increases with coarse-graining as $\sqrt{N_{\rm b}}$. Thus with the CG model MC simulations are performed, 
using an efficient particle-to-mesh calculation of non-bonded interactions which avoids neighbor lists~\cite{zhang2013}.

To parameterize the model for different $N_{\rm b}$, firstly only $\beta V_{\rm sphere}(\sigma)$ and $\beta V_{\rm bend} (\theta)$ are considered.
To match the conformational properties of ideal chains in the CG and the microscopic models $a_{\rm 1}$,  $a_{\rm 2}$, and $k_{\rm bend}$ are assigned 
the values reported by Vettorel et al.~\cite{vettorel2010}. These ensure that $P(\sigma)\sim \exp(-\beta V_{\rm sphere}(\sigma))$ and $P(\theta)\sim \exp(-\beta V_{\rm bend}(\theta))$ follow
the distributions of (a) gyration radii of $N_{\rm b}$-monomer subchains and (b) angles between imaginary lines (see Figure~\ref{method}(a))
connecting COM's of sequential $N_{\rm b}$-monomer subchains in an ideal FENE microscopic chain. Interestingly~\cite{Suter} $P(\theta)$ converges, 
increasing $N_{\rm b}$, to a universal functional form for all chains with Gaussian statistics (i.e., regardless whether they are isolated or in melt). 
Subsequently in the reference MD simulations, for $N_{\rm b}$-monomer subchains, the distributions of the distance between the COM's (only for sequential subchains) 
and of the gyration radii are calculated. Simulations of CG melts are performed, with the full CG force-field, and the remaining parameters $b_{\rm CG}$, $a_{\rm 3}$, and $\epsilon$ are 
determined iteratively so that the distributions of bond lengths and radii of soft spheres match the above reference data from the microscopic simulations. For parameter values, cf. Supporting Information.

\begin{figure}[h!]
\includegraphics[width=0.48\textwidth]{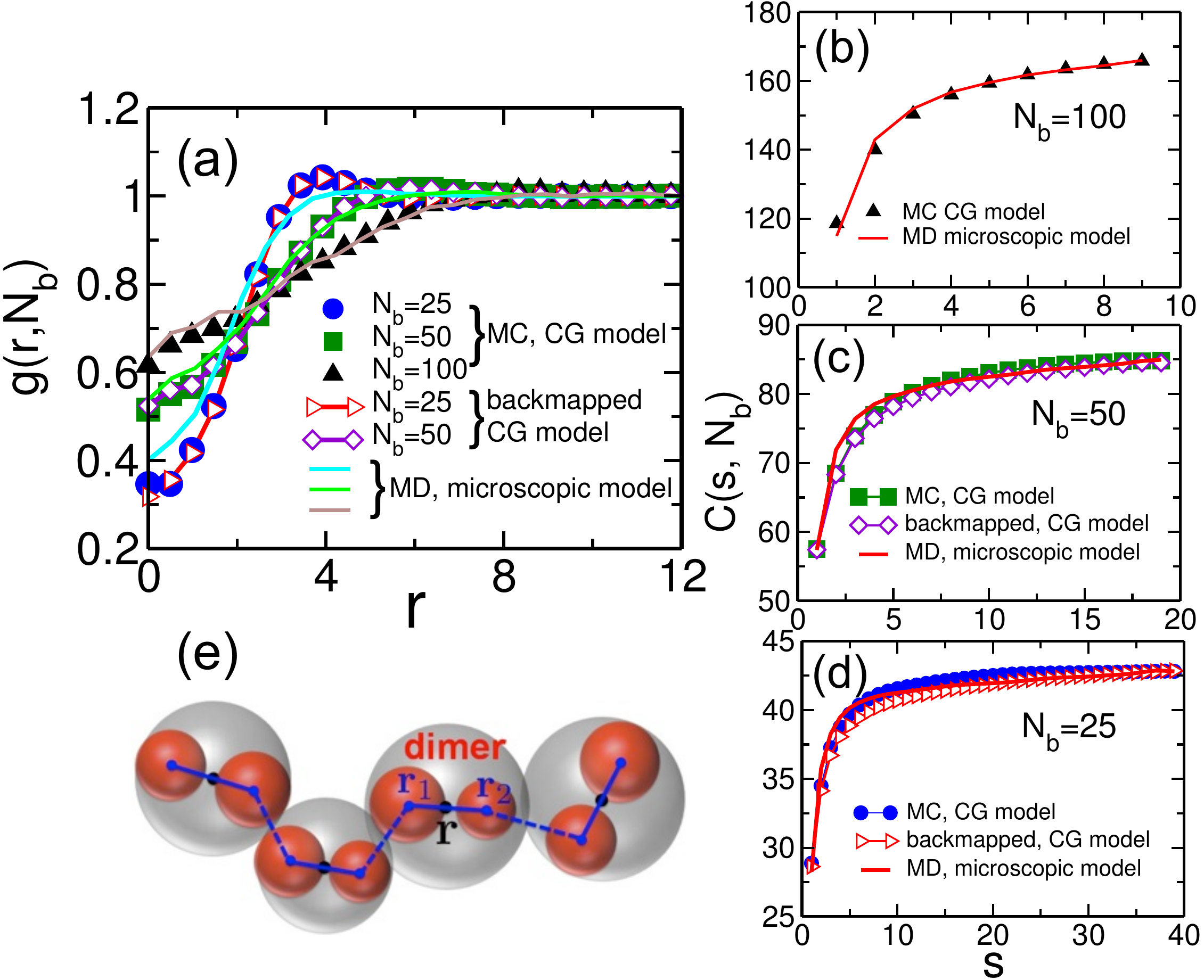}
\caption{(a) Pair distribution functions $g(r,N_{\rm b})$ of the COM's of the $N_{\rm b}$-monomer subchains in microscopic melts (solid lines)
and the centers of the soft spheres in CG melts obtained from direct MC simulations (solid symbols) and backmapping (open symbols).
(b)-(d) Internal distance plots $C(s,N_{\rm b})$ calculated from the COM's of the $N_{\rm b}$-monomer subchains
in melts described with the microscopic model (red lines) and from the centers of soft spheres in CG melts obtained from direct
MC simulations (solid symbols) and backmapping (open symbols).(e) Details of fine-graining a soft sphere chain. ``Dimer'' is a 
pair of spheres with centers located at ${\bf r}_{\rm 1}$ and ${\bf r}_{\rm 2}$, replacing a larger one with 
center located at ${\bf r}$. Solid lines show the bonds of each dimer while dashed lines denote the bonds restoring the
connectivity of the polymer after linking dimer ends.}
\label{reproducing}
\end{figure}

The CG model, parameterized as above, reproduces the remaining structural and conformational properties of the 
microscopic melt when considered at the same resolution. Figure~\ref{reproducing}(a) compares the pair correlation 
function, $g(r,N_{\rm b})$, of the COM's of the $N_{\rm b}$-monomer subchains in the microscopic melts (solid lines) with its
equivalent quantifying the packing of the centers of the soft spheres in CG melts equilibrated by MC (solid symbols). 
For resolutions $N_{\rm b} = 50$ and $100$ the two functions follow each other closely. For $N_{\rm b} = 25$ the $g(r,N_{\rm b})$ in the soft sphere model 
is somewhat more structured than its microscopic counterpart. It exhibits a deeper depletion at $r=0$, indicating that the CG non-bonded potential 
becomes harder than the effective interactions between subchains in the melt. Therefore $N_{\rm b} = 25$ corresponds approximately to the 
smallest length of the subchains in the KG model that can easily be mapped onto the current soft sphere representation.  

Figures~\ref{reproducing}(b),(c),(d) compare polymer conformations in CG melts and in reference simulations through the internal distance plot~\cite{auhl2003, wittmer2007},
$C(s,N_{\rm b}) \equiv R^2(s,N_{\rm b})/s$. For CG melts $R^2(s,N_{\rm b})$ is the mean square distance of the
centers of spheres in the same chain. In microscopic systems $R^2(s,N_{\rm b})$ is the mean square distance 
of the COM's of $N_{\rm b}$-monomer subchains in the same molecule. In both cases $s$ is the difference of the ranking numbers
of spheres (subchains) along the CG (microscopic) chain contour. Figures~\ref{reproducing}(b),(c),(d)
demonstrate that global single chain conformations are closely reproduced by the CG model. For the crude resolution, $N_{\rm b} = 100$, 
the tails ($s\geq 5$) of the two plots differ at most by $0.25\%$, while for $N_{\rm b} = 25$ the difference 
of the tails ($s\geq 25$) is less than $1\%$. More locally (smaller $s$) for both $N_{\rm b}$'s, the polymer conformations in the CG model deviate marginally 
from those in the microscopic description (the maximum difference occurs at $s=1$ or $2$ and is below $5\%$), i.e., the microscopic chains appear somewhat stiffer than the CG polymers. 
Considering the simplifications during formulation and parameterization (e.g., implementing two-body interactions and simple repulsive potentials without any 
attractive contributions~\cite{Bolhuis}) such imperfections can be expected and should not obscure the excellent performance of the model at larger length scales. More 
refined parameterizations or/and CG force fields are not considered since these small local conformational inconsistencies are easily remedied during the reinsertion 
of microscopic details. For the deviations of the internal distance plots as a function of $s$, cf. Supporting Information.

In summary, the stages of our backmapping strategy are: (a) Equilibrating a configuration at very crude resolution 
with MC simulations started from {\it random} initial configuration and sequential fine-graining until a CG melt 
with $N_{\rm b} = 25$ is generated. (b) Reinsertion of microscopic details into the last CG configuration.

Stage (a) is realized reducing at each step of the backmapping sequence the degree of coarse-graining from $N_{\rm b}$ to $N_{\rm b}/2$ (which doubles
the resolution of the model) as follows:

(1a) Firstly, each sphere is replaced with two smaller ones (i.e., a ``dimer'') as illustrated in Figure~\ref{reproducing}(e). If its center is located at ${\bf r}$,
the coordinates ${\bf r}_{\rm 1}$ and ${\bf r}_{\rm 2}$ of the centers of the two substituting smaller spheres are chosen so
that (a)  ${\bf r}_{\rm 1} + {\bf r}_{\rm 2} = 2{\bf r}$, meaning that the dimer COM coincides with the center of the replaced sphere
and (b) ${\bf r}_{\rm 1} - {\bf r}_{\rm 2} = {\bf d}$, where ${\bf d}$ stands for the vector of the bond of the dimer.
The bond orientation is chosen randomly while the length is drawn according to $\exp[-\beta V_{\rm bond}(d)]$ (the force-field parameters correspond to resolution $N_{\rm b}/2$). 

(2a) The molecular connectivity is fully restored (cf. Figure~\ref{reproducing}(e)) linking the dimers in the same chain
by the potentials $\beta V_{\rm bond}(d)$ and $\beta V_{\rm bend}(\theta)$. To relax the chain conformations only bonded interactions are considered. 
The configuration is then subjected to MC moves, displacing locally the spheres but conserving the position of the COM's of the dimers.

(3a) To obtain a packing of the reinserted spheres which is close to equilibrium, all components of the CG force-field are activated.
The radii and positions of the centers are equilibrated through MC moves (random changes of radius and local displacement) preserving the location of the COM's of the 
dimers to which the spheres belong.

(4a) All constraints are removed and the system is relaxed until all energy components reach a plateau.
The coordinates of the centers of the spheres are sampled with local displacement MC moves. This ensures that only
local relaxation occures and conformational properties on large scales are not affected.

Stage (b) comprises the following steps:

(1b) Each of the $n$ polymers in the CG melt with $N_{\rm b} = 25$ is replaced by a microscopic chain
so that its conformation complies with the overlying CG description. Namely, each of the $N_{\rm CG}$ subchains with $N_{\rm b}=25$ monomers
must fulfill two constraints: (a) the position of the COM of the subchain coincides with the location of the center
of the corresponding soft sphere, ${\bf r}_{\rm i}$, and (b) the radius of gyration (squared) of each microscopic
subchain equals the radius (squared) of the corresponding soft sphere, $\sigma_{\rm i}^2$. Therefore, we associate with each subchain
two pseudo-potentials: $\beta V_{\rm cm}= k_{\rm cm}({\bf r}_{\rm i} - {\bf R}_{\rm cm})^2$ and
$\beta V_{\rm g}= k_{\rm g}({\sigma }_{\rm i}^2 -  R_{\rm g}^2)^2$. The forces from these potentials acting on each monomer of the subchain,
combined with FENE and non-bonded intramolecular 1-2 interactions (the rest are set to zero), are employed to obtain for every CG polymer an underlying microscopic conformation
using MD. The MD is terminated once the energies $\beta V_{\rm cm}$ and $\beta V_{\rm g}$ (averaged over all subchains) reach a plateau.
This relaxation takes negligible time since, at this stage, the system presents an ensemble of non-interacting chains in external fields. 
For the values of $k_{\rm cm}$ and $k_{\rm g}$, cf. Supporting Information.

(2b) After replacing each CG polymer by an underlying microscopic chain, excluded volume is reinserted using a refined version~\cite{Livia} of the ``push-off''
procedure of Auhl et al~\cite{auhl2003}. Generally, building up microscopic monomer packing can cause significant distortion of polymer conformations with severe consequences for the structure of entanglements.
Thus non-bonded interactions are described through a ``force capped'' LJ potential, $U_{\rm LJ}(r, r_{\rm fc})$, where the original microscopic interactions are recovered as $r_{\rm fc} \rightarrow 0$. During
the ``push-off'', which is realized using MD, the deviation of the conformations from those in the equilibrium melt is minimized adjusting $r_{\rm fc}$
through a feedback loop~\cite{Livia}. This requires reference data only on local chain structure (available from small-scale  MD simulations of melts with medium sized chains).
After the ``push-off'' the original LJ potential is activated and the microscopic melt is fully equilibrated using {\it short time} MD simulations with
duration $t \simeq 2 \times 10^4 \tau$. It is emphasized that this equilibration time does {\it not} depend on chain length $N$ and is comparable to the entanglement time, $\tau_{\rm e}$, defined
as the Rouse time of a subchain with $N_{\rm e}$ monomers~\cite{Doi} (in our case~\cite{Livia} $N_{\rm e}\simeq 80$).

The backmapping strategy decomposes the equilibration of melts of long polymer chains into a sequence of steps
involving fast relaxation of short (sub)chains. The initial equilibration of properties on large scales is based on a reduced number of degrees of freedom
and soft interactions which relieve topological constraints on chain motion. Thus, if the simplest local MC moves are employed (realizing Rouse-like dynamics)
and the resolution of the start-up simulation corresponds to $N_{\rm b (o)}$, the relaxation time of the CG chains
scales as $\tau_{\rm R}\sim (N/N_{\rm b(o)})^2$. The CPU time needed to relax a system with volume $V$ should then scale
as $\tau_{\rm cg} \sim \tau_{\rm R}V\rho_{\rm o}/N_{\rm b(o)}$. The initial simulation with the crude model is the only instance when
the molecular weight of the modeled melt affects directly the equilibration time; the computational cost of the actual backmapping
procedure does {\it not} depend on chain-length. Indeed, equilibrating each level in the fine-graining hierarchy involves only dumbell motion with characteristic relaxation time $\tau_{\rm db}$,
therefore the total CPU time for equilibrating a hierarchy with $l$ levels will be $\tau_{\rm fg} \sim \tau_{\rm db} V \rho_{\rm o} \sum_{i = 1}^{l} 2^{i-1}/N_{\rm b (o)} < \tau_{\rm db} V \rho_{\rm o}\;2^{l}/N_{\rm b (o)}$.
Considering that $2^{l-1}/N_{\rm b (o)}$ is the finest resolution of the hierarchy, (i.e., a chain-length-independent quantity, which in our case 
equals $N_{\rm b} = 25$) yields $\tau_{\rm fg} \sim \tau_{\rm db} V \rho_{\rm o}$. As mentioned
above, the relaxation time of the reinsertion of microscopic details is also chain-length-independent and comparable to $\tau_{\rm e}$. 
The CPU time for accomplishing this final step will scale as $\tau_{\rm re} \sim V \rho_{\rm o} \tau_{\rm e}$.
\begin{figure}[h!]
\includegraphics[width=0.48\textwidth]{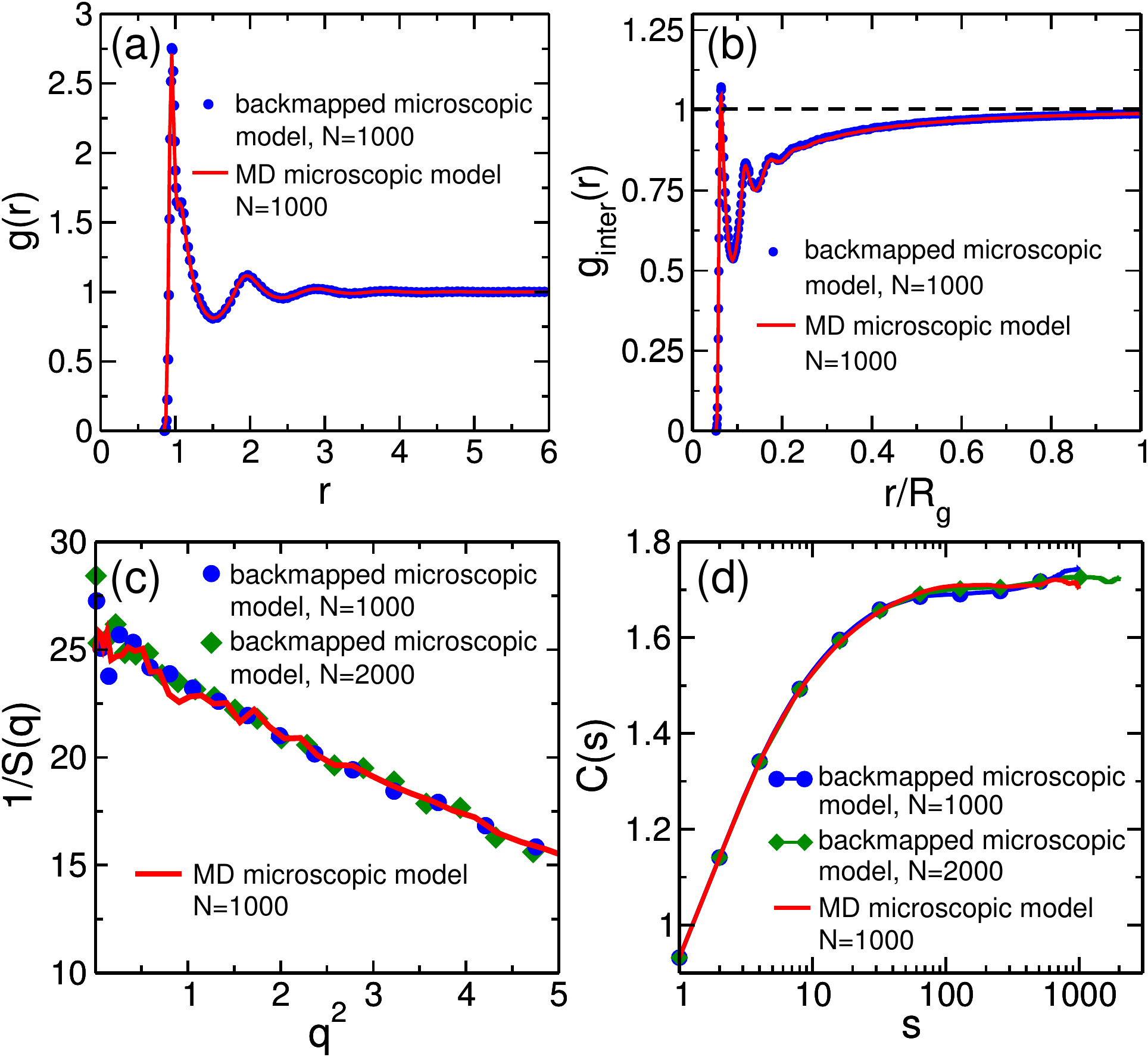}
\caption{(a) Monomer/monomer pair distribution function $g(r)$ in microscopic melts
with $N=1000$ obtained from backmapping (solid circles) and reference MD simulations (red line).
(b) Same as panel (a) but for the intermolecular monomer/monomer pair distribution function $g_{\rm inter}(r)$.
The distance, $r$, is rescaled by the average radius of gyration, $R_{\rm g}$, of the molecules. (c) and (d)
Microscopic melts with $N=1000$ and $N=2000$ (solid circles and solid rhombi) obtained from backmapping and reference MD simulations with $N=1000$ (red line)
are considered. For these systems, panel (c) shows the inverse structure factor of the density, $1/S(q)$, as a function of squared wavevector, $q^2$.
Panel (d) presents the internal distance plot, $C(s)$.}
\label{melt_micro}
\end{figure}

Here the hierarchy of the CG models incorporates three levels, $N_{\rm b} = 100$, $50$, and $25$. The agreement between
structural and conformational properties in the CG configurations created during fine-graining and those in the reference 
microscopic melts is similar to the case of direct MC simulations. For the local packing this is demonstrated calculating 
in fine-grained melts with $N_{\rm b} = 50$ and $N_{\rm b} = 25$ the pair correlation function $g(r,N_{\rm b})$ which 
is shown in Figure~\ref{reproducing}(a) with open symbols. For large $s$ the internal distance plots $C(s,N_{\rm b})$ in the 
fine-grained melts with $N_{\rm b} = 50$ and $N_{\rm b} = 25$ (Figure~\ref{reproducing}(c) and (d), open symbols) follow 
closely the $C(s,N_{\rm b})$ in the microscopic simulations demonstrating that the conformations on large scales are captured 
correctly, e.g., for $N_{\rm b} = 25$  and $s \geq 20$ the two plots differ by less than $0.1\%$. This is closer comparing to the 
direct CG simulations and we conclude that the fine-graining smoothes out some of the inaccuracies of the CG model for $N_{\rm b} = 25$. 
This can be explained considering that the correct chain structure obtained on cruder levels is transferred to the finer scales since fine-graining involves only local equilibration. 
For small $s$ the deviation of the internal distance plots in the fine-grained CG melts from the reference data reproduces the trends in direct MC simulations, i.e., it is below $5\%$. For the deviations of the internal distance plots as a function of $s$, cf. Supporting Information.

We consider now melts obtained after accomplishing the reinsertion of microscopic details into CG configurations (i.e., after stage (b)).
Figure~\ref{melt_micro}(a) presents the monomer pair correlation function, $g(r)$, calculated in systems with $N=1000$ created by
backmapping (solid symbols). It is indistinguishable from the $g(r)$ in the reference MD simulations, manifesting
that reinsertion leads to a correct microscopic structuring of the liquid. To verify that the packing of the polymer on large scales is also correctly described, 
we compare in Figure~\ref{melt_micro}(b) the intermolecular part, $g_{\rm inter}(r)$, of the pair
correlation functions in the above two systems (in each case $r$ is rescaled by the average radius of gyration, $R_{\rm g}$, of the chains).
The two plots are practically indistinguishable, confirming that the correlation hole is correctly generated (cf. the long tail behavior of the $g_{\rm inter}(r)$).
This is an important result~\cite{HoeMuel}, indicating that the correct arrangement of the polymer on scales on the order of $R_{\rm g}$ is already reproduced in the initial 
crude configuration ($N_{\rm b} =100$) generated through the direct MC simulation. This is the only stage in our scheme where chains can diffuse distances comparable or larger than $R_{\rm g}$. 
We verify that long wavelength density fluctuations are properly captured. Figure~\ref{melt_micro}(c) considers the structure factor of the density, $S(q)$, calculated for microscopic configurations of
melts with $N=1000$ and $N=2000$ created by reinsertion (solid symbols) and reference MD simulations (red line).
To highlight the behavior at small wavevectors we present the inverse structure factor, $1/S(q)$, as a function of squared wavevector, $q^2$.
It can be seen that, within the noise of the data, the results for the three systems are consistent with each other.

To confirm the agreement between the polymer conformations in melts with $N=1000$ and $N=2000$ created by backmapping and 
reference MD simulations, Figure~\ref{melt_micro}(d) presents their internal distance plots, $C(s)$, calculated on monomer 
level. $C(s)\equiv R^2(s)/s$ where $R^2(s)$ is the mean square distance of intramolecular monomers and $s$ is the difference 
of their ranking numbers along the chain contour. Indeed, after the backmapping procedure is accomplished, the relative 
deviation of the curves for all $s$ is at most $1\%$. The plots in Figure~\ref{melt_micro}(d) were obtained from a smaller number of 
independent runs than those in Figures~\ref{reproducing}(b),(c),(d), thus the data are more noisy at large $s$. 
For the deviations of the internal distance plots as a function of $s$, cf. Supporting Information.

Typical computational demands of the developed method can be illustrated considering $N=2000$ melts, created in boxes with edges equal to $5.6 R_{\rm g}$ and 
containing $n=1000$ chains, i.e., $2\times10^6$ monomers (cf. Figure~\ref{method}(c)). Probably this is the largest polymer melt that has been ever equilibrated, 
nevertheless the computational resources involved were very modest. Namely, on a single processor (2.0 GHz) the fine-graining procedure 
(stage (a)) required $8$\;h in total: $5$\;h to obtain the CG configuration at $N_{\rm b} = 100$ by MC starting from random initial configuration 
and $3$\;h for the fine-graining to reach $N_{\rm b} = 25$. The reinsertion of microscopic details until obtaining an equilibrated melt (stage (b)) required $\sim 50$\;h on 32 processors (3.0 GHz). 
The total procedure thus lasts $58$\;h. Equilibrating the same system with a configuration-assembly approach~\cite{Livia} on 32 processors requires $250$\;h. 
Here invoking only a three-level fine-graining hierarchy has as a consequence that for longer chains most of the CPU time in stage (a) is spent in the MC simulations 
generating the starting CG configuration. This CPU time can be minimized, taking full advantage of the hierarchical backmapping scheme by adding CG levels 
with cruder resolution. Thus the amount of degrees of freedom in the initial simulations can be radically reduced. Hereby, since the backmapping scheme 
itself depends only on system size and not the length of polymer chains, melts with polymerization degree up to thousand 
entanglement lengths ($N\!\sim\!10^3 \;N_{\rm e}$) can be equilibrated. 

The large samples of melts with long chains will be employed for, e.g., primitive path analysis
and calculation of rheological properties such as plateau modulus. Although we have considered as an underlying microscopic model the generic KG representation,
a similar approach can be employed for chemistry-specific models. In this case the backmapping hierarchy~\cite{VagosPS} can include an additional step transforming 
a bead-spring (or bead-rod) configuration into the atomistic representation.

We mapped polymers on chains of soft spheres with fluctuating size, interacting with simple force-fields. This description was sufficient 
to obtain microscopic configurations with correct properties after backmapping. The method can be combined with more elaborated CG potentials based 
on, e.g., integral equation theory~\cite{ClarkGuenza}. The simple link between microscopic and CG degrees of freedom is an important feature of the 
current soft sphere model, simplifying the reinsertion of the former. Facilitating reinsertion of chemical details is essential for a CG model to be 
useful for hierarchical backmapping. This should be considered when developing models with sophisticated blob 
shapes and interactions~\cite{Adamo, Murat, Eurich}, as might be necessary for studying specific chemical systems, non-linear polymers, and mixtures.

\vspace*{.2in}


\textbf{Supporting Information Available:}

Parameterization of soft sphere model for different resolutions. Deviation of internal distance plots as a function of $s$. 
This material is available free of charge via the Internet at http://pubs.acs.org.
\hspace{0pt} \\

{\bf Acknowledgment}
\\
It is a pleasure to thank Burkhard D{\"u}nweg for a careful reading of our manuscript and his helpful comments.

\end{document}